\documentclass[conference]{IEEEtran}
\usepackage[numbers]{natbib}
\usepackage{breakurl}
\usepackage[breaklinks]{hyperref}
\usepackage{graphicx}
\usepackage{tabularray}
\usepackage{amsmath}

\ifCLASSINFOpdf
\else
\fi
\begin{document}
%
\title{MRI Radiomics for IDH Genotype \\Prediction in Glioblastoma Diagnosis}

\author{
\IEEEauthorblockN{Stanislav Kozák}
\IEEEauthorblockA{\textit{Technische Hochschule Ingolstadt} \\
Email: stk9490@thi.de\\\\ \normalsize  September 23, 2024
}
}


%



%
\maketitle
\thispagestyle{plain}
\pagestyle{plain}
\begin{abstract}
Radiomics is a relatively new field which utilises automatically identified features from radiological scans. It has found a widespread application, particularly in oncology because many of the important oncological biomarkers are not visible to the naked eye. The recent advent of big data, including in medical imaging, and the development of new ML techniques brought the possibility of faster and more accurate oncological diagnosis. Furthermore, standardised mathematical feature extraction based on radiomics helps to eliminate possible radiologist bias.

This paper reviews the recent development in the oncological use of MRI radiomic features. It focuses on the identification of the isocitrate dehydrogenase (IDH) mutation status, which is an important biomarker for the diagnosis of glioblastoma
and grade IV astrocytoma.
\end{abstract}


%
\IEEEpeerreviewmaketitle

\section{Introduction}
\label{IDHMutationStatus}
Glioblastoma is the most common and aggressive type of primary brain tumour (grade IV glioma), presumably arising from neural progenitor cells. Its appearance and internal genotype are highly heterogeneous, so that biopsies taken from different parts of the tumour can be very different. \cite{MRIRadiomicsPotentialApplicationsToGlioblastoma.Hooper.2023} Non-invasive diagnosis of glioblastoma through conventional radiological methods can therefore be very difficult. 

Glioblastoma stem cells are prone to epithelial-mesenchymal transition, making them more flexible and invasive. It also contains non-malignant cells that are immunosuppressive and create a supporting microenvironment for the tumour growth \cite{MRIRadiomicsPotentialApplicationsToGlioblastoma.Hooper.2023}. These factors make the treatment very challenging, resulting in a median patient survival of less than 2 years. Recent improvements in treatment, which can include surgery, radiotherapy, chemotherapy and targeted therapy, have improved short-term survival. However, 5-year survival has remained relatively constant at 5.8\%, due to the recurrent nature of glioblastoma in most cases. As the causes are not yet well understood, there is no known way to prevent it \cite{ManagementOfGlioblastoma.2020}.

\subsection{Genotype Alterations in Grade IV Gliomas}
Grade IV gliomas can undergo different mutations and molecular alterations. In about 60\% of cases, mutations of the epidermal growth factor receptor (EGFR) have been identified, leading to a more aggressive tumour behaviour. Another important predictor of patient survival is the methylation status of 06-methylguanine DNA methyltransferase (MGMT). Methylation inhibits the production of this enzyme and slows down its DNA repair function. This may lead to a better prognosis and response to alkylating chemotherapy (usually done with temozolomide). \citep{MRIRadiomicsPotentialApplicationsToGlioblastoma.Hooper.2023, ManagementOfGlioblastoma.2020}
\subsection{IDH Mutation}
Isocitrate dehydrogenase (IDH) mutation on the chromosome 2 is one of the most important biomarkers of high-grade glioma as it significantly changes the tumour behaviour and therefore affects the survival prediction. \citep{ManagementOfGlioblastoma.2020}

A reliable method for predicting the mutation status of IDH is crucial for differentiating glioblastoma from grade IV astrocytoma and for subsequent treatment planning. \citep{GeneAlterationsSohn} IDH-wildtype (glioblastoma) is the more common and more aggressive variant, occuring predominantly in older population (median age 62 years). It develops mostly de novo, without any identifiable precursor lesion \citep{ManagementOfGlioblastoma.2020}. It is more prone to EGFR amplification (making the tumour more aggressive), but also to MGMT promoter methylation (leading to a more favourable prognosis). Subsequently, astrocytoma with IDH1 or IDH2 mutation (affecting between 5 to 13\% patients \citep{NonInvasiveCalabrese}) has overall a better prognosis and is more common in younger patients. \citep{ManagementOfGlioblastoma.2020}

Currently, the mutation status is usually determined by immunohistochemical staining with the R132H mutant IDH antibody based on tumour resection or biopsy. Non-invasive diagnosis methods based on MRI sequences continue to be explored as demonstrated in the following sections. \citep{IDHgenotypeCui}

\section{MRI Radiomics Workflow}
\begin{figure*}[h]
    \centering
    \includegraphics[width=0.75\linewidth]{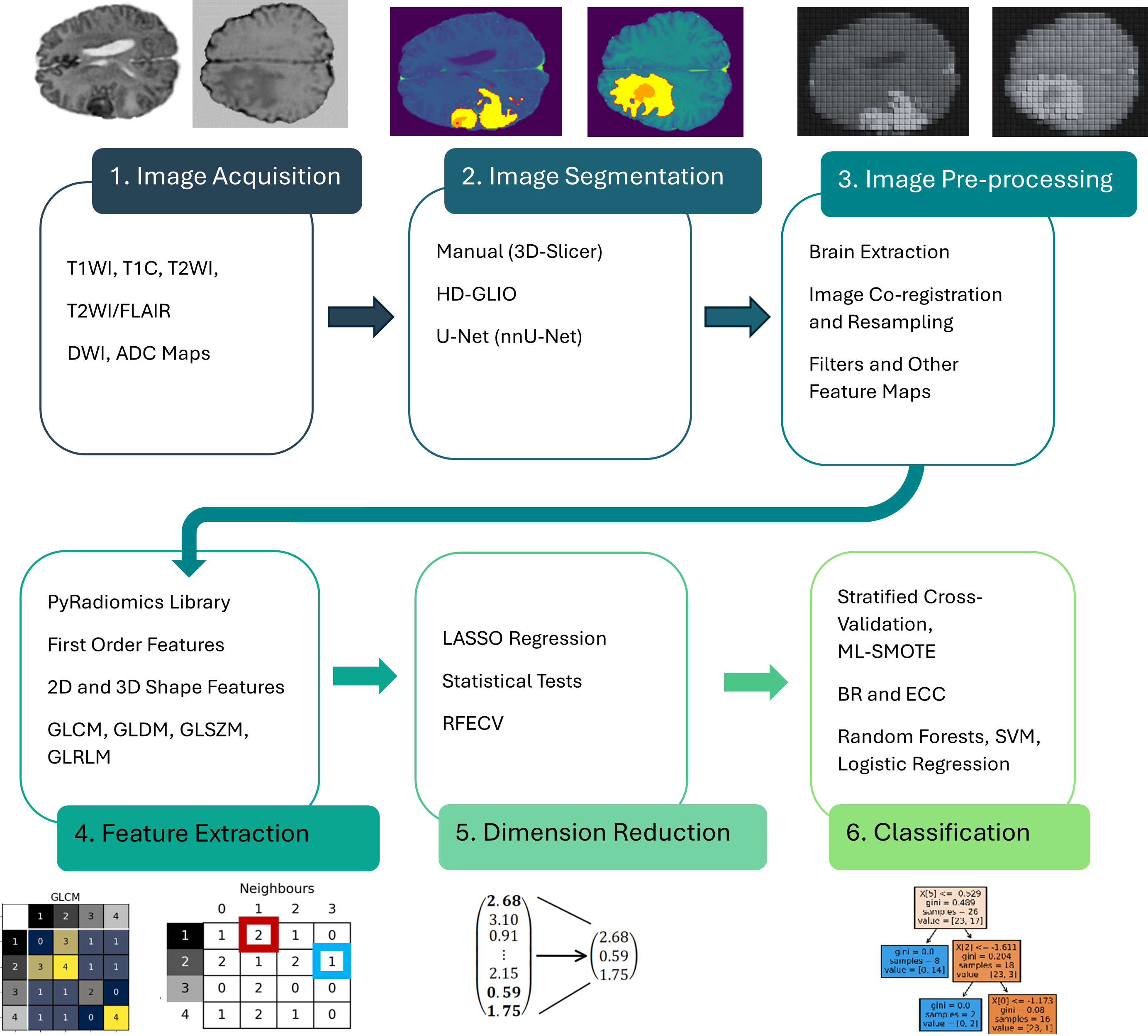}
    \caption{Radiomics pipeline based on traditional ML models.}
    \label{fig:RadiomicsWorkflowEnglish}
\end{figure*}

In this section, the commonly used radiomic pipeline for feature extraction from MRI images is introduced, as shown in figure \ref{fig:RadiomicsWorkflowEnglish}. This includes specific examples of radiomic features which are the essential components of the pipeline.

\subsection{Image Acquisition}
Although image acquisition itself is not a direct part of the radiomics pipeline, it has a major impact on the quality of the input data. The starting point is the output of magnetic resonance imaging (MRI). The images are produced by an MRI scanner emitting and measuring magnetic fields. MRI takes longer than computed tomography (CT) or positron emission tomography (PET) and is usually more expensive, but it does not require patients to be exposed to ionising radiation so it can be done more often and on patients with more critical health conditions. \citep{MRIBasicsWebsite}

An MRI scanner extracts a number of different sequences. T1-weighted (T1WI) highlights anatomical structures. A gadolinium-based contrast agent can be applied through an intravenous line to obtain a post-contrast sequence (T1C). \citep{MRIBasicsWebsite}

With modified scan settings, T2WI sequences can be acquired, which are used to detect pathological regions. Other sequences acquired by MRI or derived from the mentioned ones include T2WI/FLAIR, diffusion-weighted imaging (DWI), arterial spin labeling (ASL) or apparent diffusion coefficient (ADC) maps \citep{NonInvasiveCalabrese, IDHgenotypeCui}.

\subsection{Image Segmentation}
Segmentation is the process of delineating the region of interest (ROI) or volume of interest (VOI) from 2D or 3D images. This can be done manually, semi-automatically or automatically. Manual segmentation is the most commonly used method, but it can be time consuming, resource intensive and it can subject to observer bias. Semi-automatic methods inlude computer algorithms (e.g. region growing, tresholding) whose results are manually corrected. \citep{Radiomics-HowTo.vanTimmeren.2020}

Automatic image segmentation uses deep learning models, mostly based on the U-Net or nnU-Net architecture (for example HD-GLIO). This approach is faster and does not need expert supervision but it needs large datasets for training. The path of fully automatic segmentation seems promising but its generalisability to different datasets is still under intensive research. \citep{Radiomics-HowTo.vanTimmeren.2020}

\subsection{Image Pre-processing}
Through image pre-processing, segmented images are homogenised to provide input for feature extraction that is consistent in its characteristics. General image pre-processing includes the following procedures:
    \begin{itemize}
        \item \textit{interpolation to isotropic voxel spacing} \textit{(resampling)} – ensures the same voxel size for each sequence in the three dimensions. Because MRI yields multiple outputs, one sequence is often used as a reference scale for the rest.
        \item \textit{intensity outlier filtering} – filters out grey values outside of a predefined range.
        \item \textit{discretisation} – the scale of image intensities is divided into bins to reduce the number of possible values. \citep{Radiomics-HowTo.vanTimmeren.2020}
    \end{itemize}

For MRI, image pre-processing also includes skull stripping (brain extraction) to remove non-brain tissue which is not relevant for the analysis \citep{IDHgenotypeCui}.

Various filters or transformations can also be applied to the sequences. An open-source Python library PyRadiomics, for example, supports the following: wavelet-filters (spatial low-frequency and high-frequency filtering), Gaussian-filters (producing images with enhanced edges), square, square root, exponential and logarithmic images \citep{IDHgenotypeCui}.

\subsection{Feature Extraction}
Feature extraction uses various mathematical calculations and algorithms to quantitatively analyse the greyscale images. PyRadiomics divides extracted features into following categories: \citep{PyRadiomicsPaper.2017}
    \begin{itemize}
        \item \textit{first order (histogram based) statistics} describe the intensity distribution, for example through mean, percentiles, entropy or skewness. Entropy is calculated by the formula shown in equation \ref{eq:entropy}, with $N_g$ being the number of non-zero intensity bins, $p(i)$ the normalized histogram of intensities and $\epsilon$ an arbitrary small number:
        \begin{equation}
        \label{eq:entropy}
            entropy = -\sum^{N_g}_{i=1}p(i)*log_2(p(i)+\epsilon)
        \end{equation}
        
        \item \textit{3D shape features} are morphology descriptors, independent on the grey level intensities, computed only from a derived 3D mesh. (compactness, sphericity or surface area to volume ratio).
        \item \textit{2D shape features} are derived from a circumference mesh (perimeter, surface, elongation).
    \end{itemize}

Furthermore, PyRadiomics uses following helping matrices to extract \textit{texture-based features}: \citep{PyRadiomicsPaper.2017}
     \begin{itemize}
     \item \textit{grey level co-occurrence matrix (GLCM)} quantifies, how often each combination of grey values appears together (within a specified distance and angle).

GLCM is by default symmetrical and can be used to derive contrast, autocorrelation, joint average, difference entropy, etc. Contrast is computed by the formula \ref{eq:contrast}, iterating over the normalised GLCM entries $p(i, j)$ for each grey level combination ($i$ and $j$). The contrast is low, when the highest values of the GLCM are close to the main diagonal ($i$ and $j$ are similar):
\begin{equation}
\label{eq:contrast}
    contrast = \sum^{N_g}_{i=1}\sum^{N_g}_{j=1} (i-j)^2*p(i, j)
\end{equation}
          
          \item \textit{grey level dependence matrix (GLDM)} counts for each grey value, how often it appears with a given number of similar or equal neighbours. It is therefore used to identify clusters of similar intensities.

GLDM can be used to calculate grey level variance, high and low grey level emphasis, etc. Small and large dependence emphasis (SDE and LDE) are calculated by equations \ref{eq:sde} and \ref{eq:lde}, respectively. Here, $N_d$ is the number of neighbours and $p(i, j)$ is the normalised GLDM matrix. SDE is small if the image has small clusters (high values on the left side of the GLCM, with low number of neighbours $j$) and therefore low homogeneity. Conversely, LDE identifies large clusters and high homogeneity.
\begin{equation}
\label{eq:sde}
    SDE = \sum^{N_g}_{i=1}\sum^{N_d}_{j=1} \frac{p(i, j)}{j^2}
\end{equation}

\begin{equation}
\label{eq:lde}
    LDE = \sum^{N_g}_{i=1}\sum^{N_d}_{j=1} p(i, j) * j^2
\end{equation}

     \item \textit{grey level size zone matrix (GLSZM)} quantifying zone sizes of voxels with the same grey values (directionally independent). 
     \item \textit{grey level run length matrix (GLRLM)} describing length of equal consecutive pixels along a given angle. 
     \item \textit{neighbouring gray tone difference matrix (NGTDM)} represents the differences between grey value of a pixel and its neighbours withing a given distance.

     \end{itemize}

Further categories, such as \textit{model-based} (e.g. texture regularity analysis using autoregressive models) or \textit{transform-based} (e.g. obtained by discrete Haar wavelet transform) are also sometimes referenced. \citep{IntroductionToRadiomics.Mayerhoefer.2020}

To address the lack of reproducibility and validation of feature extraction methods in radiomic studies, the Image Biomarker Standardization Initiative (IBSI) has published guidelines and definitions for the acquisition of radiomic biomarkers \citep{IBSI.2020}. PyRadiomics also follows those guidelines.

\subsection{Feature Selection and Dimension Reduction}
Feature selection follows directly after feature extraction, as this can produce from hundred to several thousands of different features for each patient. The excessive amount of training features contains noise and can lead to overfitting of the classification model if not enough data points are available. If known, the non-reproducible features (with high intra- or inter-observer variability) should be excluded. The importance of the remaining features can be evaluated by multiple methods.

The correlation between features can be calculated to identify and remove highly correlated features. Using statistical tests, the distribution of a given feature is analysed for each class to eliminate features that are similar in all classes.

Least Absolute Shrinkage and Selection Operator (LASSO) can also be trained on normalised input features to predict the target class. It uses L1 regularisation to penalise large coefficient values, which also reduces overfitting. After fitting the model, the features with the largest respective coefficients (in absolute terms) are selected as they contribute the most to the selection.

Recursive feature elimination with cross-validation (RFECV) also identifies the important features by performing the classification itself and iteratively discarding features which do not significantly improve the accuracy.

In traditional ML research, dimension reduction (e.g. through linear discriminant analysis, principal component analysis or t-SNE) is also utilised. But because it combines multiple features together, its outputs are less interpretable, so it is not widely used in radiomic research.

Data visualisation can also help to identify important correlations. Using the tools mentioned above, the features can be grouped into correlation clusters. For each cluster, the most representative features are selected for the model fitting. This can reduce the number of features from thousands to less than 10-20. \citep{Radiomics-HowTo.vanTimmeren.2020}

\subsection{Classification Models} 
Classification or regression can be performed after the previous steps have been completed. Common classification tasks using radiomic features are prediction of the survival rate and treatment response or risk assessment, presence or stage of a particular tumour type or prediction of the tumour recurrence time. Several machine learning model types are widely used, including support vector machines, random forests, logistic regression or neural networks. Deep neural networks can only be used when larger datasets are available.

\section{Methodology}
This section provides a detailed review of the radiomic processes presented by three recent studies that focused on predicting IDH mutation status through radiomics from MRI sequences. Due to hypothesised associations between glioblastoma genotypes, the studies trained models to predict several different genetic biomarkers. These studies were selected in particular because they used different approaches to the prediction task while achieving comparable results. They all utilised Scikit-learn for the data preparation and prediction task, which is a standard Python library for traditional machine learning.  

In 2020, Calabrese et al. used a radiomic approach to predict nine different genetic biomarkers in patients with glioblastoma. From a total cohort size of 199, 195 patients were tested for the presence of IDH mutation and 190 on the MGMT promoter methylation. \citep{NonInvasiveCalabrese}

A similar study was performed by Sohn et al. in 2021. This study focused solely on predicting EGFR, IDH mutation, MGMT methylation and ATRX loss status based on a cohort of 418 patients. \citep{GeneAlterationsSohn}

Finally, Cui et al. investigated in 2023 predictive models for IDH mutation, histological phenotype (differentiation between low-grade and high-grade glioma) and Ki-67 expression level with a contrast analysis. The dataset used included 150 patients with glioblastoma and other types of glioma. \citep{IDHgenotypeCui}

\subsection{Image Acquisition}
Most of the preoperative MRI scans were performed on patients diagnosed between 2015 and 2020. All three research groups used only patient cohorts from their respective medical institutions, based on the 2016 WHO classification. Therefore, the datasets are rather small and may possibly lead to model overfitting.

Table \ref{tab:imageAcquisition} provides a detailed overview of the datasets and MRI sequence acquisition protocols. Some of the datasets were originally larger, but cases without preoperative MRI or biomarker information had to be excluded. The dataset of the third research group contained only 55 cases of diagnosed glioblastoma with a known IDH mutation status.

Due to the naturally occurring prevalence of IDH1 mutation in grade IV gliomas (between 5 and 13\% \citep{NonInvasiveCalabrese}), the classes are moderately unbalanced which needs to be compensated for in the following steps. The second group identified only 3.6\% of IDH1 mutation cases. Next-generation genetic sequencing based on biopsy or tumour resection was used to determine the IDH mutation status. Cui et al. used immunohistochemical staining with R132H mutant antibody.

Each group used different scanners and configurations. Using more scanners may make the classification task more difficult but may also help to make the dataset more generalisable. The scan produced several different sequences (4-8 in each group), including T1WI (pre-contrast and with a gadolinium-based contrast agent), T2WI, T2WI/FLAIR and DWI. Cui et al. used the DWI results to calculate ADC maps.

\begin{table}[h]
\small
\centering
\caption{Patient cohorts and image acquisition details of the reviewed studies}
\label{tab:imageAcquisition}
\begin{tblr}{
  width = \linewidth,
  colspec = {Q[200]Q[250]Q[237]Q[237]},
  hlines,
  vlines,
}
\textbf{Research group}                & \textbf{Calabrese et al. (2020)}                                                   & \textbf{Sohn et al. (2021)}                   & \textbf{Cui et al. (2023)}                            \\
\textbf{Classified biomarkers}         & IDH, ATRX, CDKN2, EGFR, MGMT, PTEN, TERT, TP53, aneuploidy of chromosomes 7 and 10 & IDH, ATRX, EGFR, MGMT                         & IDH, Ki-67-expressions, histological phenotype        \\
\textbf{Cohort size}                   & 199 patients                                                                                & 418 patients                                           & 150 patients                                                  \\
\textbf{IDH tested}  & 195 patients                                                                               & 418 patients                                          & 55 patients (GBMA), 125 (all)                                                  \\
\textbf{IDH-mutant} & 18 patients (9.2\%)                                                                         & 15 patients (3.6\%)                                    & 6 patients GBMA (10.9\%), 51 patients all (40.8\%)                                            \\
\textbf{IDH diagnosis technique}       & Next-generation genetic sequencing                                                 & Next-generation genetic sequencing            & Immuno-histochemical staining  \\
\textbf{Used scanners}                 & One 3.0 T scanner                                                                  & Two 3.0 T scanners                            & 7 scanners, mainly 1.5 T Signa HDxt \\
\textbf{Extracted MRI sequences}       & T1WI, T1C, T2WI, T2WI/FLAIR, SWI, DWI, ASL, HARDI                & T1WI, T1C, T2WI, T2WI/FLAIR & T1WI, T1C, T2WI, DWI, ADC maps      
\end{tblr}
\end{table}

\subsection{Image Segmentation}
The group of Cui et al. chose the traditional method of manual tumour segmentation. Two radiologists,  blinded to the histological and immunohistochemical results, performed  the segmentation using 3D-slicer. Each 2D slice from the T2WI sequence was segmented and then assembled to reconstruct a 3D model. \citep{IDHgenotypeCui}

The other two research groups performed automatic tumour segmentation using pre-trained deep convolutional neural networks in order to automate the entire  pipeline. Segmentation was carried out using 2D T1WI (pre and post-contrast), T2WI and T2WI/FLAIR sequences. The segmentation results were then manually examined but not corrected. \citep{NonInvasiveCalabrese, GeneAlterationsSohn}

Sohn et al. used the HD-GLIO algorithm, which separates contrast-enhancing tumour from non-enhancing T2/FLAIR signal abnormalities. The model used by Calabrese et al. consisted of three binary sub-models and segmented the images into enhancing tumour, non-enhancing tumour, surrounding tumour-related edema and background. They used Adam optimiser with learning rate decay and binary softmax cross-entropy loss for the training. \citep{NonInvasiveCalabrese, GeneAlterationsSohn}

\subsection{Image Pre-processing Techniques}
After segmentation, brain extraction was performed, for example with BET (Brain Extraction Tool) from the FMRIB Software Library (FSL) \citep{IDHgenotypeCui}. To ensure the same dimension of each sequence, the resulting images were co-registered (based on T1C or T2WI sequences) and resampled to isotropic voxel spacing (with 1x1x1 mm voxel size).

N4 bias correction with advanced normalisation tool was then applied to remove low frequency intensities caused by magnetic field inhomogeneity. Finally, the image intensities across all sequences were normalised (with $\mu=0$ and $\sigma^2=1$). \citep{GeneAlterationsSohn}

Calabrese et al. computed four additional diffusivity maps from the HARDI data: mean, axial and radial diffusivity, and fractional anisotropy, giving a total of 11 inputs for the feature extraction.

Cui et al. applied 14 filters and transformations supported by PyRadiomics to derive wavelet-filtered, Gaussian-filtered, etc. images.

\subsection{Feature Extraction}
All studies used PyRadiomics to extract radiomic features from the segmented and normalised sequences.  The respective number of features in each study is listed in table \ref{tab:featureExtraction}. The 2D and 3D shape features were extracted independent of the sequences. First order and higher order features were extracted either for each sequence \citep{GeneAlterationsSohn}, each combination of sequence and filter \citep{IDHgenotypeCui} or for each combination of sequence and segmented region (whole tumour, tumour core, 3 tumour compartments; \cite{NonInvasiveCalabrese}). This resulted in very different output sizes (660 for Sohn et al., 5300 for Calabrese et al., 6580 for Cui et al.). 

Calabrese et al. extracted all the features provided by the PyRadiomics library. The filtering criteria for shape and first-order features by Cui et al. and Sohn et al. is unknown. This lack of transparency is unfortunate as it makes the results more difficult to reproduce.

\begin{table}[h]
\small
\centering
\caption{Extracted features}
\label{tab:featureExtraction}
\begin{tblr}{
  width = \linewidth,
  colspec = {Q[329]Q[244]Q[194]Q[175]},
  hlines,
  vlines,
}
\textbf{Research group}                  & \textbf{Calabrese et al. (2020)} & \textbf{Sohn et al. (2021)} & \textbf{Cui et al. (2023)} \\
\textbf{Extracted from}             & 11 sequences, 5 segmented regions (55 combinations)                               & 4 sequences                          & 5 sequences, 14 image types                         \\
\textbf{Shape features}                  & 26 per segmentation                               & Not specified               & 14 per sequence                         \\
\textbf{First order features}            & 19 per combination                               & Not specified               & 18 per image type                         \\
\textbf{Higher order (texture) features} & 75 per combination                               & Not specified               & 75 per image type                         \\
\textbf{Total features per patient}      & \textbf{$26 * 5 + (19 + 75) * 55 = \boldsymbol{5300}$}                    & \textbf{660}                & {$(14 + (18+75)*14)*5=\boldsymbol{6580}$}              
\end{tblr}
\end{table}

\subsection{Feature Selection}
Sohn et al. and Cui et al. used LASSO regression to discard features with a low predictive value. 

Cui et al. also performed Mann-Whitney U test for each class, selecting only input features whose distributions varied. They then calculated the Pearson correlation coefficient between the remaining features and removed highly correlated features. 

Calabrese et al. and Cui et al. used RFECV to select the final features. In each iteration step, RFECV trained a simple classification model to quantify their importance. The least important features were then removed and the process was repeated several times for different dataset splits to avoid overfitting.

\subsection{Classification Models}
The low natural prevalence of the IDH mutation makes the data set unbalanced. To prevent the model from underperforming on the minority class, Calabrese et al. used 10-fold stratified cross validation (with 60/40 train-test split), which ensures the same class distribution in each set. Cui et al. trained the classifier for the data set with all gliomas which was not imbalanced (40.8\% patients with IDH mutation). They used repeated k-fold cross validation (80/20 split repeated 30 times).

Another method is synthetic data generation which was used by Sohn et al. They run the Multi-Label SMOTE algorithm to generate new data points of minority classes while maintaining the associations between the labels. They used a 70/30 ratio for the train test split. \citep{GeneAlterationsSohn}

There are two main approaches to multi-label classification: binary relevance (BR) and classifier chain (CC). BR uses separate independent binary classifiers trained on each label. In CC, each label is also predicted by a binary model, but it can use the result of previous labels and therefore takes into account the correlation between different labels. When the optimal order of classification is not known, an ensemble classifier chain (ECC)  can be used, which iterates over chains with different classifier orders. Calabrese et al. and Cui et al. chose the binary relevance approach, while Sohn et al. compared both approaches. \citep{GeneAlterationsSohn}

The choice of model and loss function can have a significant impact on the training time and on the accuracy. Linear models are usually faster to train but they learn less representative decision boundary than non-linear models.

Calabrese et al. treated each label prediction as a binary regression task, predicting the probabilities for both the negative and positive class. They used a random forest regressor and randomised search for hyperparameter tuning. \citep{NonInvasiveCalabrese}

Sohn et al. used a linear kernel support vector machine trained by SGD with manual hyperparameter tuning. They also tried out 10 different classifier orders for the ECC approach, evaluated by the mean absolute Shapley values. The optimal classifier order found was IDH-ATRX-MGMT-EGFR. 

Cui et al. built independent classifiers for each MRI sequence. They applied support vector machines and logistic regression with 9 different loss functions, including hinge, logarithmic, Huber or epsilon insensitive. They also tried L1 and L2-regularisation and their combination (Elastic Net). Based on the results of the single sequence classification, they built final classifiers combining T1C and ADC sequences because those two sequences reached the most accurate predictions. \citep{IDHgenotypeCui}

\section{Evaluation and Discussion}
\label{ResultsChapter}
In this chapter, a comparison and discussion of the results and conclusions of the three studies under review will be presented. This includes a comparison of the model performance, the optimal radiomic features found, and the associations between the biomarkers themselves. 

The following sections conclude the paper by describing the potential and limitations of predicting IDH mutation status based on radiomic features in clinical practice.

\subsection{Result Comparison}
Table \ref{tab:results} compares the results obtained by the final classification and regression models for the IDH mutation prediction. It is important to note that Cui et al. predicted the IDH mutation status for all gliomas, not just grade IV. Based on the ROC curve, they probably also considered IDH-wild-type as a positive class, unlike the other studies. This discrepancy may be due to the fact that the positive class usually indicates more rare and dangerous cases at the same time. In other words, Cui et al. chose the more dangerous class as positive, whereas the other studies chose the more rare class as positive. To make it easier to compare the results between studies, the metrics using the IDH mutation as a positive class were also included (shown in brackets).

The high recall values indicate the ability of the model to detect most cases of IDH mutation (minimising false negatives and the type II error). This is offset by more false positives and therefore lower precision (and higher type I error). Optimising for recall (type II error) is helpful for the clinical practice because it avoids higher costs and exposing patients to riskier treatment, as the diagnosis is rarely overestimated. However, it results in more cases being underestimated.

Labelling the minority class as positive also validates the usage of precision, recall and their harmonic mean (F1-Score) as metrics. However, MCC and AUC are often more preferable metrics for imbalanced data sets as they are symmetric under class labelling.

Very good predictive performance for distinguishing glioblastoma from IDH-mutated astrocytoma was achieved by Calabrese et al. (MCC 0.62). Cui et al. achieved an even better performance (MCC 0.68) for classifying IDH mutation and IDH wild-type, but this could be caused by including lower grade gliomas in the data set. \citep{NonInvasiveCalabrese, IDHgenotypeCui}

\begin{table}[h]
\centering
\caption{Final model evaluation}
\label{tab:results}
\begin{tblr}{
  width = \linewidth,
  colspec = {Q[144]Q[317]Q[252]Q[225]},
  hlines,
  vlines,
}
\textbf{Research}  & \textbf{Calabrese et al. (2020) } & \textbf{Sohn et al. (2021)} & \textbf{Cui et al. (2023), all gliomas} \\
\textbf{Precision} & 0.50                              & 0.26                        & 0.93 (0.73)                      \\
\textbf{Recall}    & 0.93                              & 1.0                         & 0.81 (0.89)                      \\
\textbf{F1-Score}  & 0.62                              & 0.42                        & 0.87 (0.80)                      \\
\textbf{MCC}       & 0.62                              & 0.48                        & 0.68 (0.68)                      \\
\textbf{AUC}       & 0.95                              & 0.96                        & 0.88 (0.88)                     
\end{tblr}
\end{table}

\subsection{Found Features and Biomarker Associations}
Surprisingly, the features with the highest predictive value differed significantly between the three studies. There was little to no overlap in the imaging sequences (except for T1C), segmented regions (tumour core, contrast enhancing tumour, etc.) or the feature groups (first order, shape features, etc.) in the top selected features. 

Features extracted from T1C sequences were among the most efficient. Calabrese et al. identified variance on the whole tumour segmented region, Sohn et al. identified coarseness (NGTDM feature), surface area to volume ratio (shape feature) and maximum correlation coefficient (GLCM feature) on contrast-enhancing segmented regions as important. Cui et al. did not rank the features by the predictive performance, but only listed the 21 features selected for the final classifier.  They identified several first-order features and features from GLRLM, GLSZM or GLCM on logarithmic or wavelet-filtered images from both T1C and ADC sequences.

For Calabrese et al, diffusivity metrics had also a high predictive value. One of those was high grey level emphasis, a feature based on GLDM that quantifies the diffusivity of the non-enhancing tumour produced by the DWI sequence. Another was the kurtosis of tumour-related edema from the mean diffusivity mask.

The optimal classifier order found by Sohn et al. highlights a significant effect of IDH prediction on ATRX and MGMT classification (as mentioned in chapter \ref{IDHMutationStatus}): \textit{"IDH mutation increases the overall genomic CpG methylation and is strongly associated with MGMT promoter methylation."} \citep{GeneAlterationsSohn}

The found correlation of IDH and MGMT prediction is particularly helpful as MGMT alone was difficult to predict, for example in Calabrese et al. also found that ATRX mutations are more common in IDH mutant gliomas but rare in IDH wild-type.

In conclusion, the studies reviewed all achieved satisfactory classification performance, but did not find any fully overlapping radiomic features that would allow unambiguous identification of the IDH mutation. This is due to the different segmentation, image processing and feature extraction techniques used.

It is also possible that the models were highly dependent on the small and unbalanced datasets, as the performance was worse on independent validation data. 

\subsection{Potential and Limitations of Radiomics for IDH Genotype Prediction}
The presented radiomic approach may contribute to a non-invasive and faster diagnosis when differentiation between glioblastoma and IDH-mutated astrocytoma is required for further treatment planning. The development of a highly accurate classification model could validate or replace biopsy or radiologist diagnosis.

Further research is needed to address several issues. Most importantly, the procedures and techniques used in the radiomic pipeline should be standardised and consistently documented. The efforts of the IBSI to standardise feature extraction have been effective as there were only small differences between the reviewed studies in the extracted features (all have used standard features by PyRadiomics).

The second issue is the generalisability of the classification model. Larger datasets with different patient demographics, health conditions and scanner types are needed to develop a reliable classifier. Furthermore, IDH2 mutation could be included as it is also a indicator of grade IV astrocytoma (albeit with a lower prevalence).

This also raises the question of the overall suitability of the radiomic approach. When larger datasets are available, the traditional machine learning classifiers could be replaced by a deep learning network. The whole pipeline could also be replaced by an end-to-end framework, for example using a convolutional neural network for classification. In this case, the interpretability of the model's diagnosis would suffer. However, as has been shown, the optimal radiomics features found also convey characteristics that are not visible to humans and they vary significantly between studies, so there is no ultimate feature combination that could unambiguously solve the classification task while being fully interpretable.





%

\end{document}